\documentclass[epj]{svjour}
\RequirePackage{graphicx} 
\usepackage{latexsym}
\usepackage{graphics}
\usepackage{enumitem}
\usepackage{cite} 
\usepackage{amsmath}
\usepackage{hyperref}
\usepackage{cite}
\usepackage{amsmath,amssymb}
\usepackage{xcolor}

\begin{document}
\title{Anisotropic stars made of exotic matter within the complexity factor formalism}
\author{
\'Angel Rinc{\'o}n \inst{1} 
\thanks{E-mail: \href{mailto:angel.rincon@ua.es}{\nolinkurl{angel.rincon@ua.es}} }
\and 
Grigoris Panotopoulos \inst{2} 
\thanks{E-mail: \href{mailto:grigorios.panotopoulos@ufrontera.cl}{\nolinkurl{grigorios.panotopoulos@ufrontera.cl}} }
\and 
Il{\'i}dio Lopes \inst{3}
\thanks{E-mail: \href{mailto:ilidio.lopes@tecnico.ulisboa.pt}{\nolinkurl{ilidio.lopes@tecnico.ulisboa.pt}} }
}                     
\offprints{}          
\institute{
Departamento de Física Aplicada, Universidad de Alicante, Campus de San Vicente del Raspeig, 
E-03690 Alicante, Spain.
\and 
Departamento de Ciencias F\'isicas, Universidad de la Frontera,
Casilla 54-D, 4811186 Temuco, Chile.
\and 
Centro de Astrof{\'i}sica e Gravita{\c c}{\~a}o-CENTRA, Instituto Superior T{\'e}cnico-IST, Universidade 
de Lisboa-UL, Av.~Rovisco Pais, 1049-001 Lisboa, Portugal.
}
\date{Received: date / Revised version: date}
%
\abstract{
We investigate exotic stars composed of dark energy within the context of Einstein's General Relativity, by applying an extended Chaplygin gas equation-of-state. To account for anisotropies, we utilize a formalism based on the complexity factor to obtain numerical solutions.
By applying well-established criteria, we demonstrate that the solutions are physically valid and well-behaved. In addition, a comparison with a more conventional approach is also conducted.
\PACS{
      {PACS-key}{discribing text of that key}   \and
      {PACS-key}{discribing text of that key}
     } 
} 
\maketitle

\section{Introduction}

Any reasonable modern cosmological model must include Dark Energy (DE).   
Nevertheless, the nature and origin of Dark Energy remain a mystery despite its fundamental importance in modern theoretical cosmology 
\cite{SupernovaSearchTeam:1998fmf,SupernovaCosmologyProject:1998vns,Freedman:2003ys}.  
As it is well known, a cosmological model made of only matter and radiation cannot lead to accelerated solutions to the universe as predicted by Einstein’s Theory of General  Relativity (GR)
 \cite{Einstein:1916vd}. This kind of solution is obtained by including a constant $\Lambda$ in Einstein's field equations \cite{Einstein:1917ce}, i.e., by adding the contribution of the dark energy.  Despite its simplicity, such accelerated cosmological model is in exceptional agreement with a vast amount of observational data. Such a cosmological model is known as the concordance cosmological model or the $\Lambda$CDM model.
 Nevertheless, $\Lambda$  suffers from the cosmological constant ongoing problem \cite{Weinberg:1988cp,Zeldovich:1967gd}.  	
 Additionally, this $\Lambda$--problem is amplified by the current values estimation of the Hubble constant $H_0$, using high red-shift CMB data and local measurements at low red-shift data, e.g.,
 \cite{Ryden:2017dxw,Colin:2019opb,Verde:2013wza,Bolejko:2017fos}.  In fact, the value of the $H_0$ computed by the PLANCK Collaboration \cite{Planck:2015fie,Planck:2018vyg}, $H_0 =
 (67-68)~km/(Mpc \: sec)$, is lower than the value estimated from local measurements \cite{Riess:2016jrr,Riess:2018byc}, $H_0 = (73-74)~km/(Mpc \: sec)$. This $H_0$ tension points to a cosmological model with new physics \cite{Mortsell:2018mfj,Kazantzidis:2018rnb,Gannouji:2018ncm,Alvarez:2020xmk}.

\smallskip

Over the years, this incomplete picture of the cosmological concordance model has motivated the arrival of many new and alternative models. We can classify recent DE cosmological models into two generic categories: (i) alternative theories of gravity for which the solutions have additional corrective terms compared to the standard case; (ii) by employing a new dynamical degree of freedom by means of  a convenient equation-of-state.  In the first class of models, one  finds, for instance, Scalar-Tensor theories of gravity \cite{Brans:1961sx,Brans:1962zz,Sanchez:2010ng,Panotopoulos:2017clp}, brane-world models \cite{Randall:1999ee,Randall:1999vf,Dvali:2000hr,Langlois:2002bb,Maartens:2003tw}  and  f(R) theories of gravity \cite{Sotiriou:2008rp,DeFelice:2010aj,Hu:2007nk,Starobinsky:2007hu}; and for the second class, one finds models such as  k-essence \cite{Armendariz-Picon:2000ulo}, phantom  \cite{Arefeva:2004odl},
quintessence  \cite{Ratra:1987rm}, quintom \cite{Lazkoz:2006pa}, or tachyonic  \cite{Bagla:2002yn}.   For a good review article on the dynamics of dark energy see for instance \cite{Copeland:2006wr}.

\smallskip

In this work, we will focus our study on the generalized Chaplying gas equation-of-state \cite{Szydlowski:2020ilx}, widely used in many cosmological model extensions. Here, we study the properties of relativistic astrophysical objects, where we opt to use the same equation of state.

\smallskip

In studies of compact relativistic astrophysical objects the authors usually focus on stars made of an isotropic fluid,
where the radial pressure $P_r$ equals the tangential pressure $P_\bot$. However, celestial bodies are not always made of isotropic fluid only. In fact under certain conditions the fluid can become anisotropic. The review article of Ruderman \cite{Ruderman:1972aj} mentioned for the first time such a possibility: this author makes the observation that relativistic particle interactions in a very dense nuclear matter medium could lead to the formation of anisotropies. The study on anisotropies in relativistic stars has received a boost by the subsequent work of \cite{Bowers:1974tgi}.
Interestingly,  Ivanov \cite{2010IJTP...49.1236I} has shown that by considering a compact object to be an anisotropic star, the effects of shear, electromagnetic field, etc, can be automatically taken into account.
 Indeed, anisotropies can arise in many scenarios
of a dense matter medium, like phase transitions \cite{Sokolov:1998er}, pion condensation \cite{Sawyer:1972cq}, or in presence of type 3A super-fluid \cite{Kippenhahn:2012qhp}.
See also \cite{MakANDHarko,Deb:2016lvi,Deb:2015vda} for more recent works on the topic, and references therein. In these works relativistic models of
anisotropic quark stars were studied, and the energy conditions were fulfilled. In particular, in \cite{MakANDHarko} an exact analytical solution was obtained, in \cite{Deb:2016lvi} an attempt was made to find a singularity free solution to Einstein’s field equations,
and in \cite{Deb:2015vda} the Homotopy Perturbation Method was employed, which is a tool that facilitates to tackle Einstein’s field
equations. What is more, alternative approaches have been considered to incorporate anisotropies into known isotropic solutions \cite{Gabbanelli:2018bhs,Ovalle:2017fgl,Ovalle:2017wqi}.

\smallskip

Beyond the collisionless dark matter paradigm, self-interacting dark matter has been proposed as an attractive
solution to the dark matter crisis at galactic scales \cite{Tulin:2017ara}. In this scenario one can imagine relativistic stars made entirely of self-interacting dark matter, see e.g. \cite{Li:2012sg,Maselli:2017vfi,Panotopoulos:2018enj}. In a similar way, given that the current cosmic acceleration
calls for dark energy, very recently a couple of works appeared in the literature, where the authors entertain the
possibility that stars made of dark energy or more generically exotic matter just might exist \cite{NewtonSingh:2020rsk,Tello-Ortiz:2020svg}. 

\smallskip

These exotic stars are unique objects like any other compact object that manifest themselves across many multi-messenger signals like gravitational waves,  neutrinos, cosmic rays and electromagnetic radiation from radio up to gamma-rays.
For instance, we will be able to test many of these stellar models using the data from the present and next generation of gravitational wave detectors such as LIGO, Virgo, KAGRA and LISA. 	

\smallskip

In the present work, we propose to study non-rotating dark energy stars with anisotropic matter assuming a generalized equation-of-state of the form $p = -B^2/\rho + A^2\rho$ (with $A$ and $B$ being constants). A simplified version of this, known as a Chaplygin equation-of-state, was introduced in Cosmology long time ago to unify the description of non-relativistic matter and the cosmological constant \cite{Kamenshchik:2001cp,Bento:2002ps,Debnath:2004cd}. Such a generic equation-of-state is originated by a viscose matter, that when considered in a cosmological context gives rise to
the unification of dark matter and dark energy \cite{Szydlowski:2020ilx}.

\section{Relativistic spheres within GR}

We will consider a static, spherically symmetric object (static fluid), and we will assume locally certain anisotropy, bounded by a spherical surface $\Sigma$.  The line element considering  Schwarzschild--like coordinates is written as
\begin{equation}
ds^2=e^{\nu} dt^2 - e^{\lambda} dr^2 -
r^2 d\Omega^2,
\label{metric}
\end{equation}
where $\nu(r)$ and $\lambda(r)$ are, as always, the corresponding metric potential, depending on the radial coordinate only,  and $d\Omega^2\equiv \left( d\theta^2 + \sin^2\theta d\phi^2 \right)$ correspond to the element of solid angle. We
will take: $x^0=t; \, x^1=r; \, x^2=\theta; \, x^3=\phi$. 
The classical Einstein field equations with a vanishing cosmological constant are given by 
\begin{equation}
	G^\nu_\mu=8\pi G T^\nu_\mu,
	\label{Efeq}
\end{equation}	
where $G$  is Newton's constant ($G=1$)  and $T^\nu_\mu$ is the energy-momentum tensor. In the comoving frame, the matter content is described by an anisotropic fluid with energy density $\rho$, radial pressure $P_r$, and tangential pressure $P_\bot$.
The covariant energy-momentum tensor can be expressed in local Minkowski coordinates as $T^{\mu}_{\nu} =\{ \rho , P_r, P_\bot, P_\bot\}$, and the resulting field equations take the form:

\begin{eqnarray}
\rho &=& -\frac{1}{8\pi}\left[-\frac{1}{r^2}+e^{-\lambda}
\left(\frac{1}{r^2}-\frac{\lambda'}{r} \right)\right],
\label{fieq00}
\\
P_r &=& -\frac{1}{8\pi}\left[\frac{1}{r^2} - e^{-\lambda}
\left(\frac{1}{r^2}+\frac{\nu'}{r}\right)\right],
\label{fieq11}
\\
P_\bot &=& \frac{1}{32\pi}e^{-\lambda}
\left(2\nu''+\nu'^2 -
\lambda'\nu' + 2\frac{\nu' - \lambda'}{r}\right),
\label{fieq2233}
\end{eqnarray}
where the derivatives with respect
to $r$ are denoted by  primes.

As it is well known, we can combine the last equations to produce the hydrostatic equilibrium  equation (also known as the  generalized Tolman-Opphenheimer-Volkoff equation), i.e.,
\begin{equation}
-\frac{1}{2}\nu'\left( \rho + P_r\right)-P'_r+\frac{2}{r}\left(P_\bot-P_r\right)=0.\label{Prp}
\end{equation}
We can express the equilibrium equation as a balance between three forces, namely the gravitational force ($F_g$), hydrostatic force ($F_r$), and anisotropic
force ($F_p$), which we define as follows for convenience:
 \begin{equation}
 F_g=-\frac{\nu'\left( \rho+ P_r\right)}{2},\;
F_r= -P'_r\; {\rm and}\;
F_p=\frac{2\Pi}{r}.
\label{Forces}
 \end{equation}  
where $\Delta \equiv \Pi=P_\bot-P_r$.
Thus, equation (\ref{Prp}) can now be expressed as:
\begin{equation}
F_g+F_r+F_p=0.\label{Frp}
\end{equation}	
The equilibrium for a compact star is maintained by the balance of three forces, as established by the previous equation (\ref{Frp}) \cite{Prasad:2021eju}. It is noteworthy that if $F_p$ is equal to zero, then the standard TOV equation is obtained. When $P_\bot>P_r$ (or $\Pi>0$), $F_p>0$ generates a repulsive force in equation (\ref{Frp}) that counteracts the attractive forces of $F_g$ and $F_r$. Conversely, when $P_\bot<P_r$ (or $\Pi<0$), $F_p<0$ becomes an additional attractive force that acts in conjunction with the other forces.
Alternatively, we can remove the $\nu'$-dependence
in equation (\ref{Prp})
 to obtain a more convenient equation, namely
\begin{equation}
P'_r=-\frac{(m + 4 \pi P_r r^3)}{r \left(r - 2m\right)}\left( \rho + P_r\right)+\frac{2}{r}\left(P_\bot-P_r\right),\label{ntov}
\end{equation}
To do that, we have used the relation
\begin{equation}
\frac{1}{2}\nu' =  \frac{m + 4 \pi P_r r^3}{r \left(r - 2m\right)},
\label{nuprii}
\end{equation}
Furthermore, the mass function $m$ is obtained by:
\begin{equation}
R^3_{232}=1-e^{-\lambda}=\frac{2m}{r},
\label{rieman}
\end{equation}
or, 
\begin{equation}
m = 4\pi \int^{r}_{0} \tilde r^2\rho \  d\tilde r.
\label{m}
\end{equation}
Now, let us rewrite the energy-momentum tensor as follow
\begin{equation}
T^{\mu}_{\nu}=\rho u^{\mu}u_{\nu}-  P
h^{\mu}_{\nu}+\Pi ^{\mu}_{\nu},
\label{24'}
\end{equation}
Firstly, we set the four-velocity as 
$u^{\mu} = (e^{-\frac{\nu}{2}},0,0,0)$, and the four acceleration, $a^\alpha=u^\alpha_{;\beta}u^\beta$, whose any non--vanishing component is $a_1 = -\nu
^{\prime}/2$. Subsequently, the set $\{ \Pi^{\mu}_{\nu}, \Pi, h^{\mu}_{\nu}, s^{\mu}, P \}$ is taken according to
\begin{eqnarray}
\Pi^{\mu}_{\nu} &=& \Pi\bigg(s^{\mu}s_{\nu}+\frac{1}{3}h^{\mu}_{\nu}\bigg)
\\
\Pi &=& P_{\bot}-P_r \label{Delta}
\\
h^\mu_\nu &=& \delta^\mu_\nu-u^\mu u_\nu
\\
s^{\mu} &=& (0,e^{-\frac{\lambda}{2}},0,0)\label{ese}
\\
P & \equiv & \frac{1}{3}\Bigl( P_{r}+2P_{\bot} \Bigl)
\end{eqnarray}
with the  properties
$s^{\mu}u_{\mu}=0$,
$s^{\mu}s_{\mu}=-1$.
To obtain the exterior solution, we match the problem with the Schwarzs\-child spacetime, as follows:
\begin{equation}
ds^2= \left(1-\frac{2M}{r}\right) dt^2 - \left(1-\frac{2M}{r}\right)^{-1} dr^2 -
r^2  d\Omega^2.
\label{Vaidya}
\end{equation}
The problem should be supplemented using certain boundary conditions on the surface
$r=\text{constant}=R$, with $R$ being the radius of the star. Therefore, we require that the first and second fundamental forms are continuous across that surface. This condition implies that:
\begin{eqnarray}
e^{\nu_\Sigma} &=& 1-\frac{2M}{R},
\label{enusigma}
\\
e^{-\lambda_\Sigma} &=& 1-\frac{2M}{R},
\label{elambdasigma}
\\
\left[P_r\right]_\Sigma &=& 0.
\label{PQ}
\end{eqnarray}
Here, the subscript $\Sigma$ indicates that the quantity is evaluated on the boundary surface $\Sigma$. In conclusion, it is worth noting that the last three equations are both necessary and sufficient conditions for a smooth matching of the two metrics (\ref{metric}) and (\ref{Vaidya}) on the surface $\Sigma$.

\section{Anisotropic matter: Complexity factor}

In what follows, we will briefly summarize the underlying physics behind the definition of the complexity factor, focusing on the astrophysical relevance of such quantity. Let us first start mentioning the seminal paper by Herrera \cite{Herrera:2018bww}, where a new and non-trivial way to reveal when static self-gravitating objects are anisotropic was properly introduced.
Furthermore, this revised definition aims to address two issues that were identified in earlier attempts to define complexity.
The first problem appears when the probability distribution (which appear in the definition of ``disequilibrium” and information) is replaced by the energy density of the fluid distribution \cite{Sanudo:2008bu}.
The second issue arises from the recognition that previous definitions of complexity only take into account the energy density of the fluid, while neglecting other crucial components such as pressure.
Thus, the new definition introduced by L.H. try to make progress by fixing the above mentioned issues.

\smallskip

Originally, the new definition of the complexity factor was investigated only under a mathematical point of view (see \cite{Sharif:2018pgq,Sharif:2018efi,Abbas:2018cha,Herrera:2019cbx} and references therein). However, the real value of such definition becomes evident when we use it as a supplementary condition to close the set of differential equations of a self-gravitational system.
Additionally, the complexity factor may serve as a self-consistent method for integrating anisotropies \cite{Arias:2022qrm,Andrade:2021flq}, which has been explored in recent studies \cite{Maurya:2022yva,Maurya:2022cyv,Sharif:2022akn,Sharif:2022sad,Govender:2022ome,Bogadi:2022yqb,Bargueno:2022yob,Sadiq:2022uwj} and their associated references.


As was previously pointed out, the complexity factor appears in the orthogonal splitting of the Riemann tensor for static self-gravitating fluids with spherical symmetry, and for a detailed step-by-step computation, we should see the original paper \cite{Herrera:2018bww} and also \cite{Gomez-Lobo:2007mbg}.
Therefore, while we will refrain from delving deeply into the orthogonal decomposition of the Riemann tensor, we must still establish the following quantities:
\begin{eqnarray}
Y_{\alpha \beta} &=& R_{\alpha \gamma \beta \delta}u^{\gamma}u^{\delta}, \label{electric} 
\\    
Z_{\alpha \beta} &=& R^{*}_{\alpha \gamma \beta
\delta}u^{\gamma}u^{\delta} = \frac{1}{2}\eta_{\alpha \gamma
\epsilon \mu} R^{\epsilon \mu}_{\quad \beta \delta} u^{\gamma}
u^{\delta}, \label{magnetic} 
\\
X_{\alpha \beta} &=& R^{*}_{\alpha \gamma \beta \delta}u^{\gamma}u^{\delta}=
\frac{1}{2}\eta_{\alpha \gamma}^{\quad \epsilon \mu} R^{*}_{\epsilon
\mu \beta \delta} u^{\gamma}
u^{\delta}, \label{magneticbis}
\end{eqnarray}
Please, notice that  the symbol $*$ represent the dual tensor, namely
\begin{equation}
   R^{*}_{\alpha \beta \gamma \delta}=\frac{1}{2}\eta_{\epsilon \mu \gamma \delta}R_{\alpha \beta}^{\quad \epsilon \mu} 
\end{equation}
and $\eta_{\epsilon \mu \gamma \delta}$ is the well-known Levi-Civita tensor.
Taking advantage of the decomposition of the Riemann tensor, we rewrite the set of scalars
$\{Y_{\alpha \beta}, Z_{\alpha \beta}, X_{\alpha \beta}\}$ in term of the physical variables, i.e., %
\begin{eqnarray}
Y_{\alpha\beta} &=& \frac{4\pi}{3}(\rho +3
P)h_{\alpha\beta}+4\pi \Pi_{\alpha\beta}+E_{\alpha\beta},\label{Y}
\\
Z_{\alpha\beta} &=& 0,\label{Z}
\\
X_{\alpha\beta} &=& \frac{8\pi}{3} \rho
h_{\alpha\beta}+4\pi
 \Pi_{\alpha\beta}-E_{\alpha\beta}.\label{X}
\end{eqnarray}
Notice that the corresponding tensor $E_{\alpha \beta}$ (defined as $E_{\alpha \beta}=C_{\alpha \gamma \beta
\delta}u^{\gamma}u^{\delta}$) is given by
\begin{equation}
E_{\alpha \beta}=E \bigg(s_\alpha s_\beta+\frac{1}{3}h_{\alpha \beta}\bigg),
\label{52bisx}
\end{equation}
with
\begin{equation}
E=-\frac{e^{-\lambda}}{4}\left[ \nu ^{\prime \prime} + \frac{{\nu
^{\prime}}^2-\lambda ^{\prime} \nu ^{\prime}}{2} -  \frac{\nu
^{\prime}-\lambda ^{\prime}}{r}+\frac{2(1-e^{\lambda})}{r^2}\right],
\label{defE}.
\end{equation}
The following properties must be satisfied:
 \begin{eqnarray}
 E^\alpha_{\,\,\alpha}=0,\quad E_{\alpha\gamma}=
 E_{(\alpha\gamma)},\quad E_{\alpha\gamma}u^\gamma=0.
  \label{propE}
 \end{eqnarray} 
Even more, as was also demonstrated by \cite{Herrera:2009zp}, the tensors $\{ Y_{\alpha\beta}, Z_{\alpha\beta}, X_{\alpha\beta} \}$ can be represented in term of alternative scalar functions. 
Considering the tensors $X_{\alpha \beta}$ and $Y_{\alpha \beta}$ in the static case, the so-called structure scalars  $X_T, X_{TF}, Y_T, Y_{TF}$ can be written in term of the physical variables as follow:
\begin{eqnarray}
X_T     &=& 8\pi  \rho ,  \label{esnIII} \\
X_{TF}  &=& \frac{4\pi}{r^3} \int^r_0{\tilde r^3 \rho ' d\tilde r} \label{defXTFbis}, \\
Y_T     &=& 4\pi( \rho  + 3 P_r-2\Pi) \label{esnV}, \\
Y_{TF}  &=& 8\pi \Pi- \frac{4\pi}{r^3} \int^r_0{\tilde r^3 \rho' d\tilde r} \label{defYTFbis}.
\end{eqnarray}
From Eqs.\eqref{defXTFbis}-\eqref{defYTFbis}, the local anisotropy of pressure is  determined by  $X_{TF}$ and $Y_{TF}$ via the following relation:
\begin{equation}
8\pi \Pi = X_{TF} + Y_{TF} \label{defanisxy}.
\end{equation}
When the complexity vanishes ($Y_{TF}=0$), it implies the following relation between the energy density and the aniso\-tropic factor:
	\begin{equation}
		\Pi(r) = \frac{1}{2r^3} \: \int^r_0{\tilde r^3 \rho'(\tilde r) d\tilde r}.
\end{equation}
The last condition has also been significantly investigated along years introducing, via alternative ansatzs, several concrete forms of the anisotropy $\Pi \equiv P_{\perp} - P_r $ and different equations of state (see for instance \cite{Panotopoulos:2018joc,Panotopoulos:2018ipq,Moraes:2021lhh,Gabbanelli:2018bhs,Panotopoulos:2019wsy,Lopes:2019psm,Panotopoulos:2019zxv,Abellan:2020jjl,Panotopoulos:2020zqa,Bhar:2020ukr,Panotopoulos:2020kgl,Panotopoulos:2021obe,Panotopoulos:2021dtu} and references therein).
Given that a profound comprehension of the idea of complexity is still under construction, the connection between $\Pi$ (or more precisely, any equation of state $P_{\bot} \equiv P_{\bot}(\rho)$) and the definition of complexity factor $Y_{TF}$ is still missing.


\section{Discussion}
This work examines anisotropic stars composed of exotic matter through the well-established complexity formalism. Our study focuses on numerically computing solutions for a realistic compact distribution of matter, and comparing these results with those obtained using the conventional formalism within the framework of GR. To this end, we employ a generalized Chaplyin equation-of-state to close the system. In the figures presented, we show the evolution of several key quantities of the star. Notably, we observe that:
\begin{enumerate}[label=\roman*)]
\item 	
The mass function increases, while the anisotropic factor, energy density, and pressures decrease throughout the star.
\item
The speed of sound, both radial and tangential, increases and decreases, respectively, with both values lower than $c_0^2 \equiv 1$, the relativistic adiabatic index, while $\Gamma(r)$ increases and remains greater than $\Gamma_0 \equiv 4/3$.
\item
Furthermore, the corresponding energy conditions are satisfied. 
\end{enumerate}
Therefore, based on these numerical results, we can confidently assert that the complexity factor formalism is a robust approach for obtaining well-defined solutions within the context of compact stars.

\smallskip

As a supplementary check, we obtained interior solutions using a more standard approach. This involved adding external constraints to close the system of differential equations, and we employed numerical methods to carry out the calculations. As a toy model, we considered an anisotropic factor, $\Pi(r)$, defined by the equation	
	\begin{equation}
		\Pi(r) = - \bigg(\frac{r}{a}\bigg)^2  \rho(r),
	\end{equation}
	where $a$ is a dimensionful parameter with units of length. This parameter encodes the strength of the anisotropy.
	
The anisotropic factor used in \cite{Moraes:2021lhh} has a simple mathematical form that satisfies fundamental requirements: it has the correct dimensions, is negative, and vanishes at the star's center ($\Pi(r = 0) =0$). To further explore its behavior, we investigate two scenarios characterized by large ($a_{\text{large}} = 30~\text{km}$) and small ($a_{\text{small}} = 10~\text{km}$) values of $a$.

Although it is not necessary, one can derive the differential equation
	\begin{equation}
		\Pi'(r) = \frac{2}{r}  \Pi(r) - \frac{r^2}{a^2}  \rho'(r),
	\end{equation}
	from the anisotropic factor ansatz given above. This equation bears a striking resemblance to the differential equation
	\begin{equation}
		\Pi'(r) = - \frac{3}{r}  \Pi(r) + \frac{\rho'(r)}{2},
	\end{equation}
which is obtained using equation (36) and the condition that the complexity vanishes, i.e., $Y_{TF}=0$. Equation (41) can be straightforwardly derived as follows: We first take the derivative of both sides of equation (39) with respect to $r$ and obtain
\begin{equation}
	\Pi'(r) = \frac{1}{a^2} [ 2 r \rho + r^2  \rho'(r) ].
\end{equation}
We then use the anisotropic factor definition again,
\begin{equation}
	\frac{\Pi(r)}{r} = - \frac{r}{a^2}  \rho(r).
\end{equation}

Our main finding is that when the normalized anisotropy, $\Pi(r)/B$, is comparable to the one studied within the complexity factor formalism, the resulting solution violates causality, as shown in Fig. 4 for the small $a$ case. On the other hand, if the solution is realistic and meets all the criteria, the star's anisotropy is much lower, as illustrated in Fig. 5 for the large $a$ case, while retaining a similar mass.

\smallskip
We illustrate the mass-to-radius profiles  for three models in Fig. 6. The left panel shows the models within the complexity factor formalism, while the right panel presents a more standard approach for the case of large $a$. The three models are:
\begin{equation}
A = \sqrt{0.4}, \; \; \; \; \; B = 0.23 \times 10^{-3} km^{-2}
\end{equation}
for Model 1,
\begin{equation}
A = \sqrt{0.425}, \; \; \; \; \; B = 0.215 \times 10^{-3} km^{-2}
\end{equation}
for Model 2, and
\begin{equation}
A = \sqrt{0.45}, \; \; \; \; \; B = 0.2 \times 10^{-3} km^{-2}
\end{equation}
for Model 3.

\smallskip

The curves show that the star's radius first reaches a maximum value, followed by the maximum mass of the star. The complexity factor formalism predicts smaller and lighter objects compared to the conventional approach, despite both cases having a negative anisotropic factor.
The complexity factor is an effective method to include modifications in the structure of compact stars resulting from anisotropy. This leads to alterations in the TOV equations for such stars, balancing three forces: gravity, hydrostatic and anisotropic forces.

\smallskip
In the final figure (Fig.~7), we present the mass-to-radius relationships for isotropic and anisotropic stars using both approaches for Model 1. We begin by plotting the M-R profile for stars composed of isotropic matter (dashed line), and then explore anisotropies within the vanishing complexity formalism, yielding the cyan curve.
Turning to the conventional method, we can examine the impact of the continuous parameter characterizing the ansatz for the anisotropic factor on the profiles, which are shown in the other three curves in the figure. As the anisotropy increases, the profile gradually shifts towards the one corresponding to complexity. However, at a certain point, the violation of causality occurs, marking the upper limit for the anisotropic factor using the conventional method. This critical point signifies where the solution becomes unrealistic or unviable, and it is crucial to halt the analysis to ensure the solution's physical validity. As a result, the last permitted profile differs significantly from the one obtained using complexity.
 

\begin{figure*}[ht!]
\centering
\includegraphics[width=0.32\textwidth]{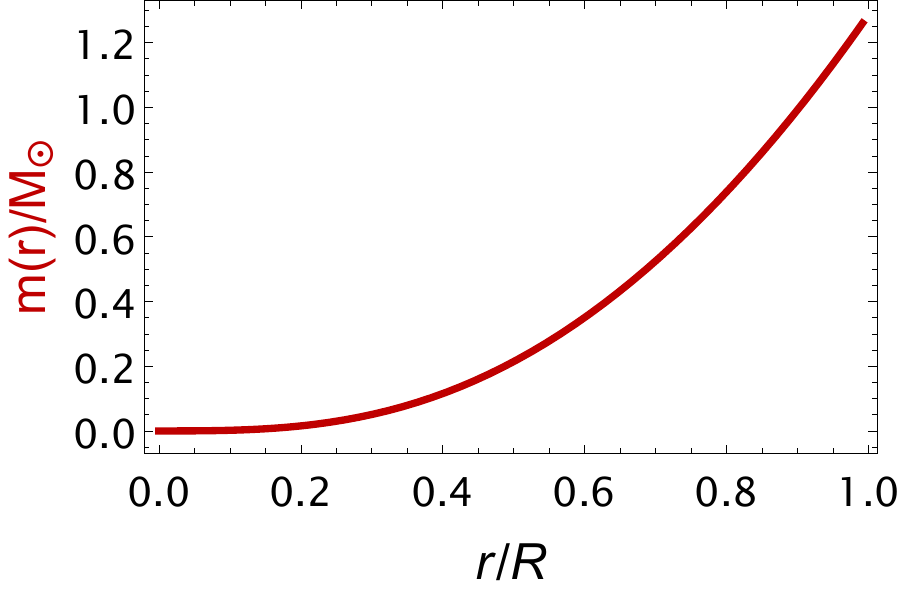} \
\includegraphics[width=0.32\textwidth]{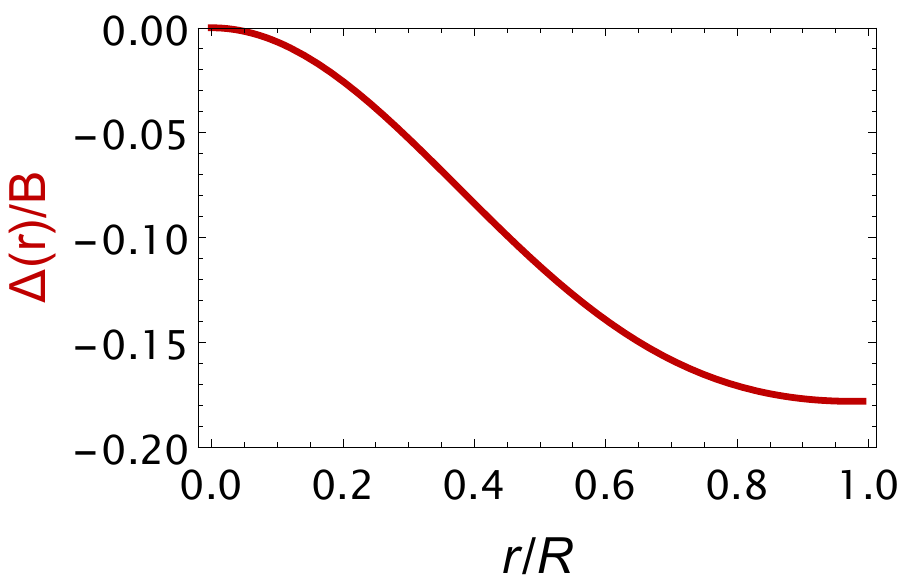}  \
\includegraphics[width=0.32\textwidth]{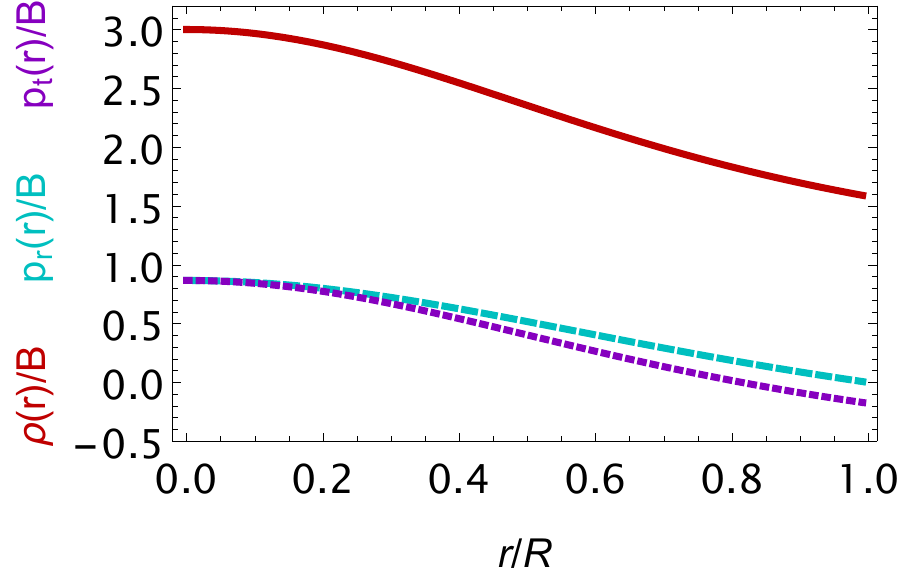} 
\caption{	
The left panel shows the mass function in solar masses, the middle panel displays the anisotropic factor, and the right panel illustrates the energy density and pressures as a function of the radial coordinate throughout the star, specifically for anisotropic DE stars within the complexity factor.
}
\label{fig:1} 	
\end{figure*}


\begin{figure*}[ht!]
\centering
\includegraphics[width=0.48\textwidth]{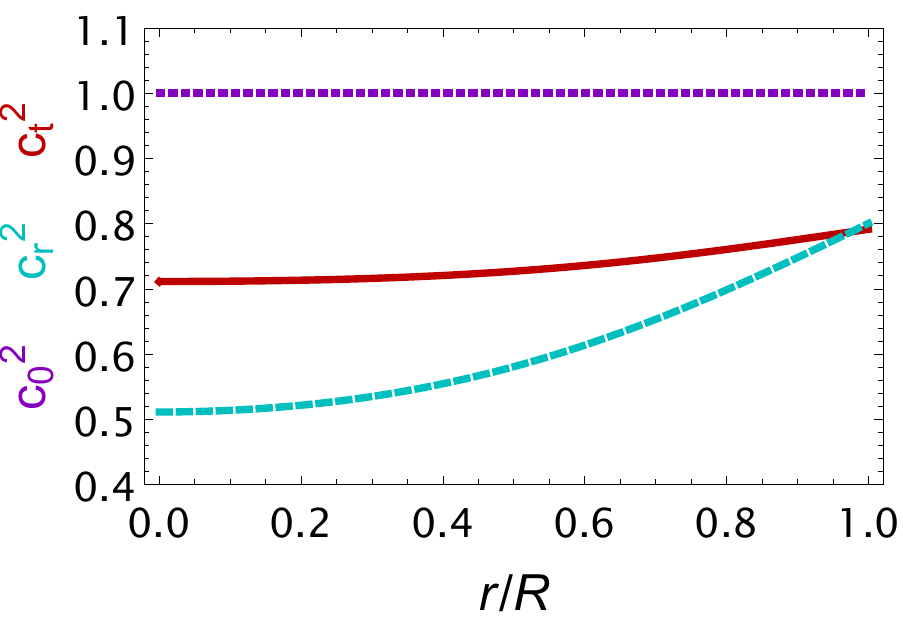} \
\includegraphics[width=0.48\textwidth]{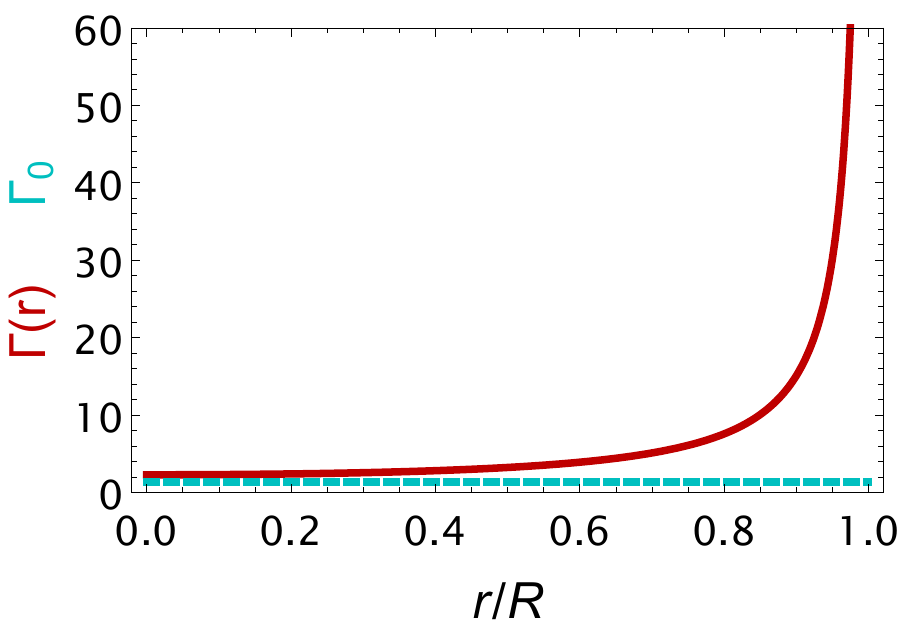} 
\caption{
The left panel displays the speed of sounds, while the right panel exhibits the relativistic adiabatic index as a function of the radial coordinate throughout the anisotropic DE stars.	
}
\label{fig:2} 	
\end{figure*}


\begin{figure*}[ht!]
\centering
\includegraphics[width=0.48\textwidth]{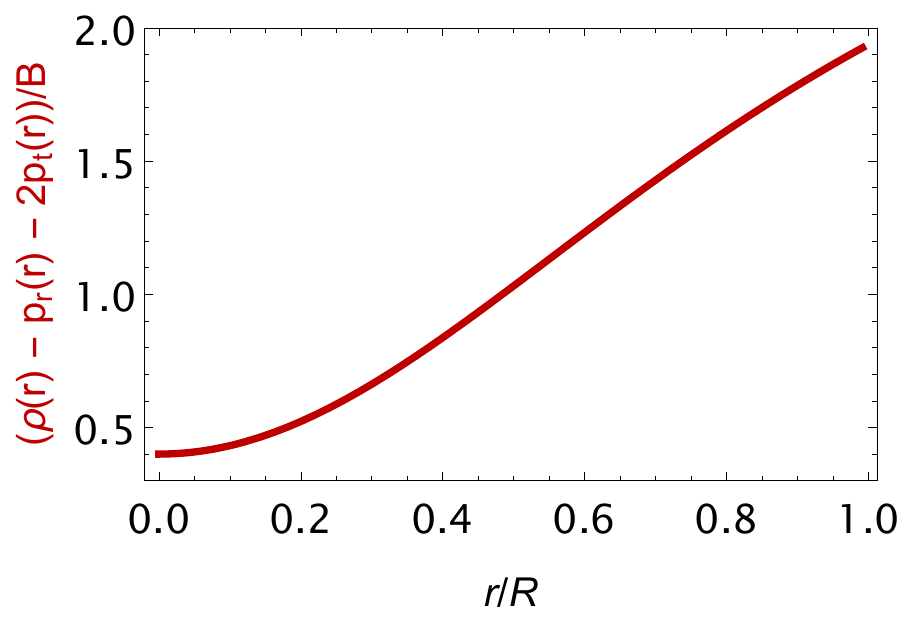} \
\includegraphics[width=0.48\textwidth]{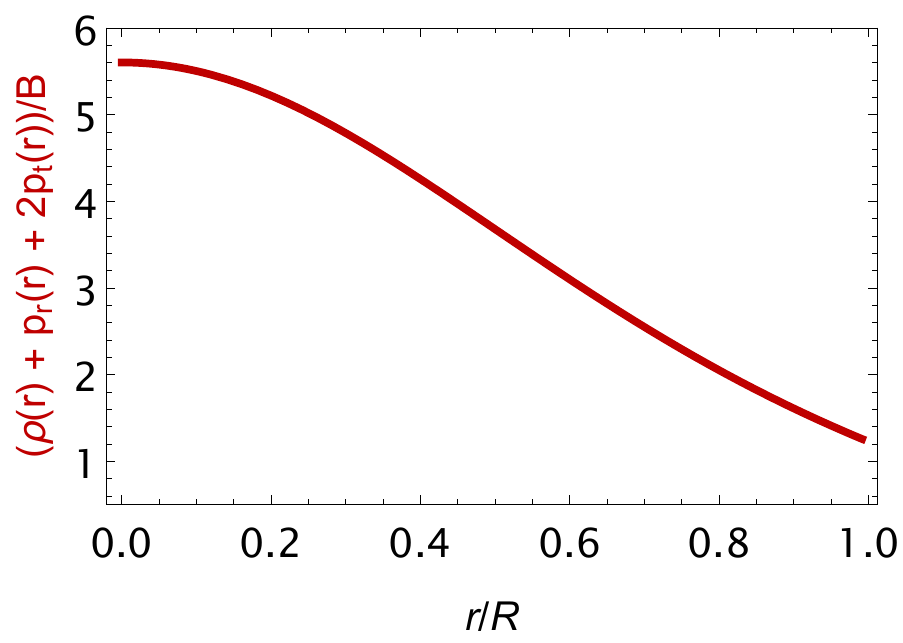} 
\caption{
The energy conditions for anisotropic DE stars within the complexity factor are presented as a function of the radial coordinate throughout the star.	
}
\label{fig:3} 	
\end{figure*}


\begin{figure*}[ht!]
\centering
\includegraphics[width=0.32\textwidth]{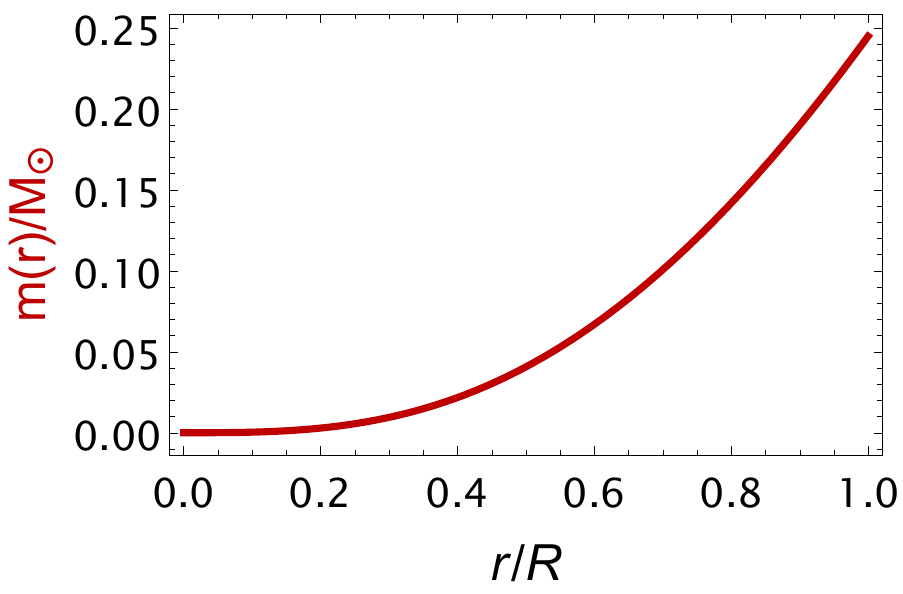} \
\includegraphics[width=0.32\textwidth]{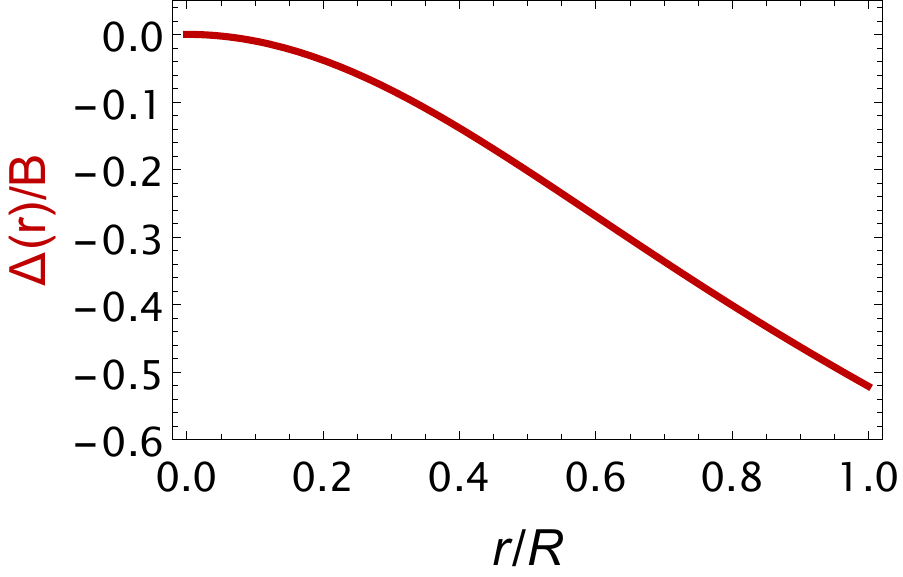} \
\includegraphics[width=0.32\textwidth]{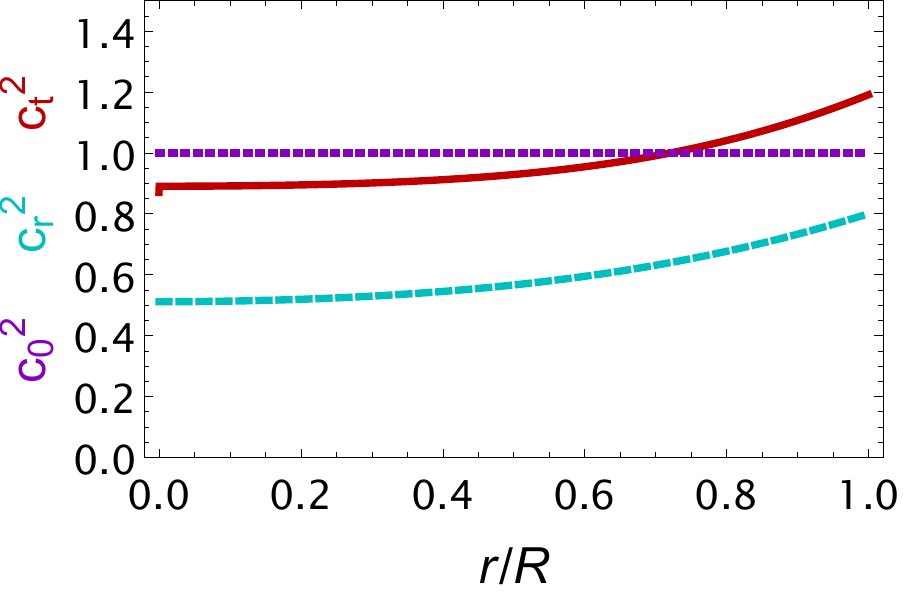} 
\caption{
Anisotropic DE stars using a  standard approach and assuming a small value for $a$: the left panel of the figure shows the mass function in solar masses, the middle panel shows the anisotropic factor, and the right panel shows the sound speeds, all plotted as a function of the radial coordinate throughout the star.
}		
\label{fig:4} 	
\end{figure*}


\begin{figure*}[ht!]
\centering
\includegraphics[width=0.32\textwidth]{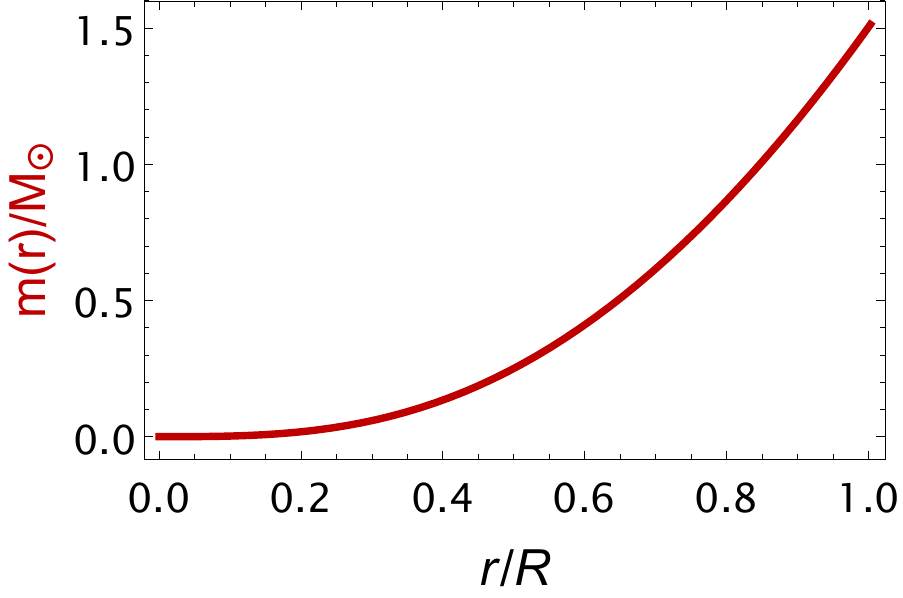} \
\includegraphics[width=0.32\textwidth]{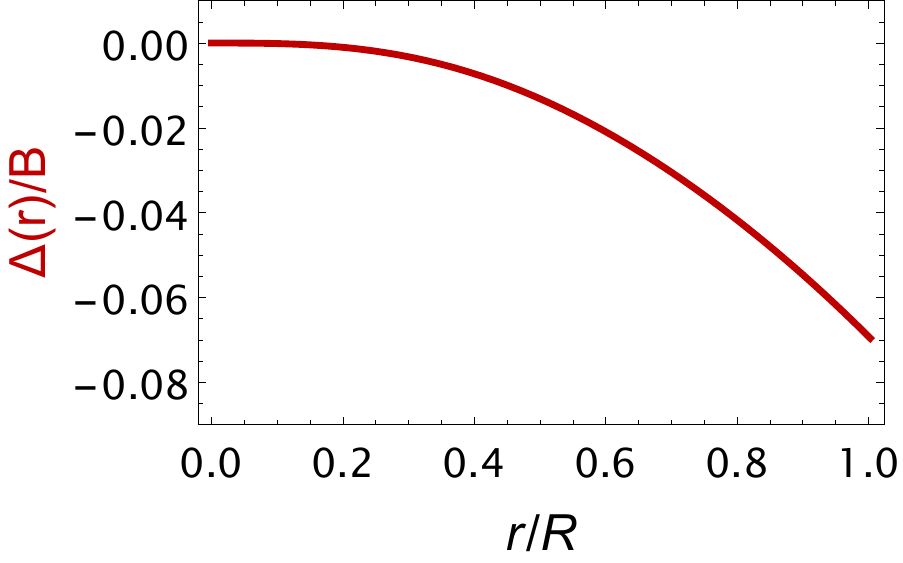} \
\includegraphics[width=0.32\textwidth]{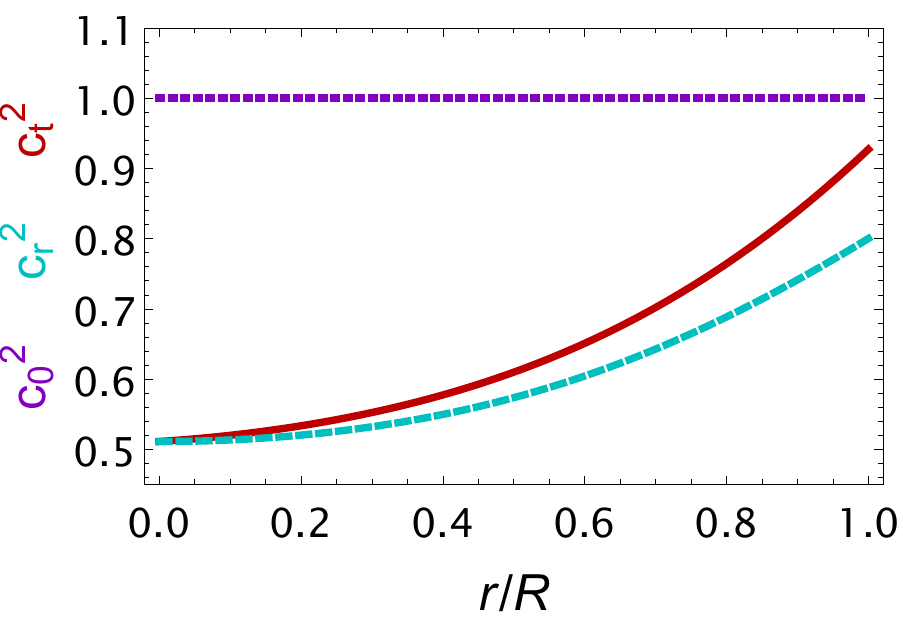} \caption{
Anisotropic DE stars using a standard approach and assuming a large value for $a$: the figure displays the mass function in solar masses in the left panel, the anisotropic factor in the middle panel, and the sound speed in the right panel, all plotted as a function of the radial coordinate throughout the star.
}
\label{fig:5} 	
\end{figure*}

\begin{figure*}[ht!]
\centering
\includegraphics[width=0.48\textwidth]{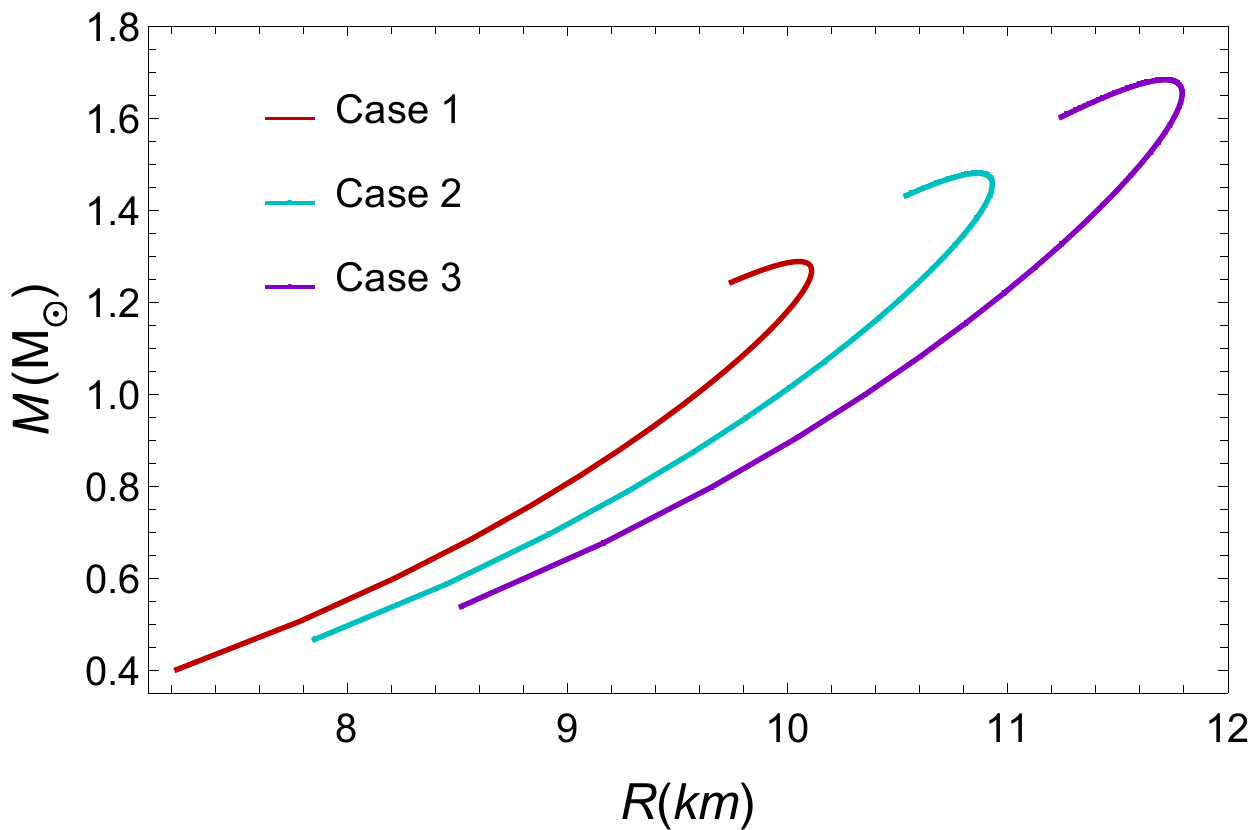} \
\includegraphics[width=0.48\textwidth]{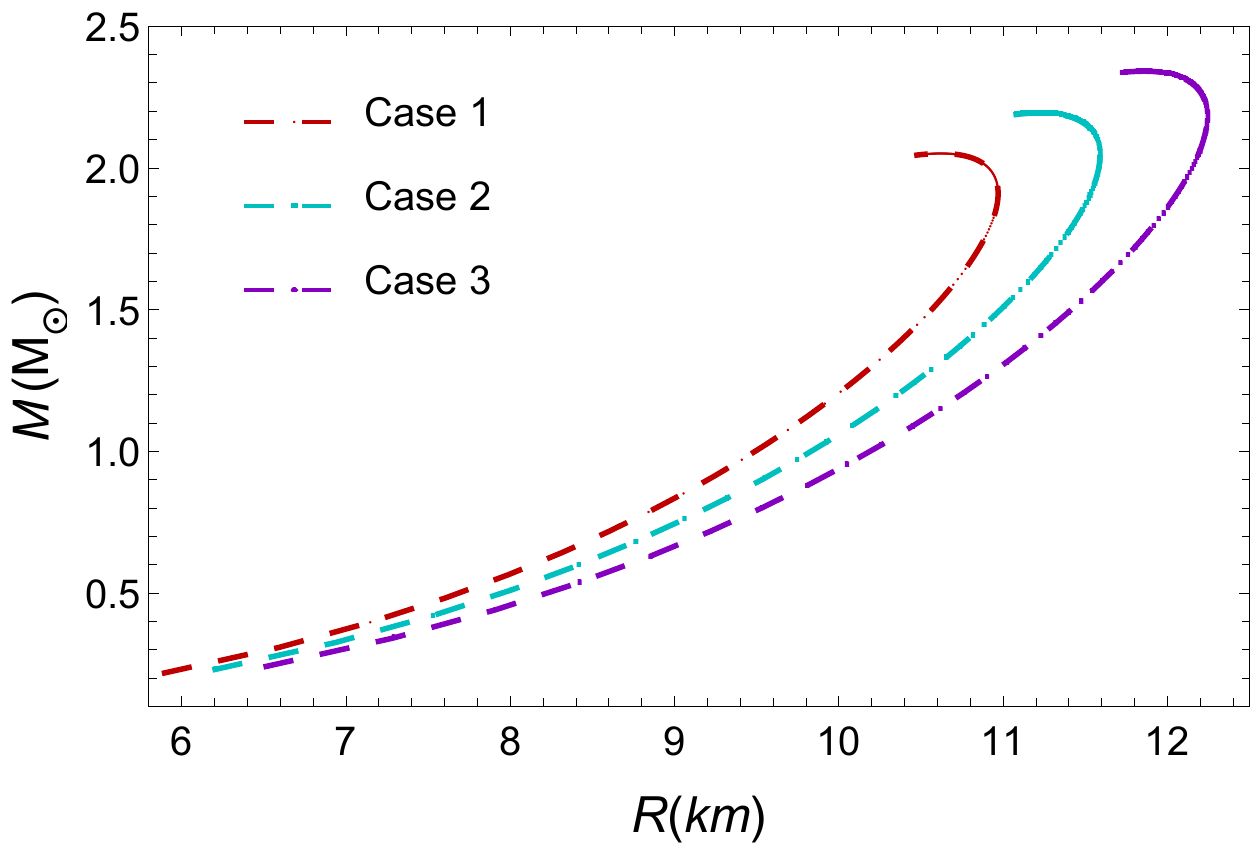} 
\caption{
The  mass-to-radius profiles for two alternative approaches: the left panel shows the profiles for three models corresponding to the curves following the complexity formalism, while the right panel shows the profiles for three models corresponding to curves obtained in the conventional scenario.
}
\label{fig:6} 	
\end{figure*}

\begin{figure*}[ht!]
\centering
\includegraphics[width=0.75\textwidth]{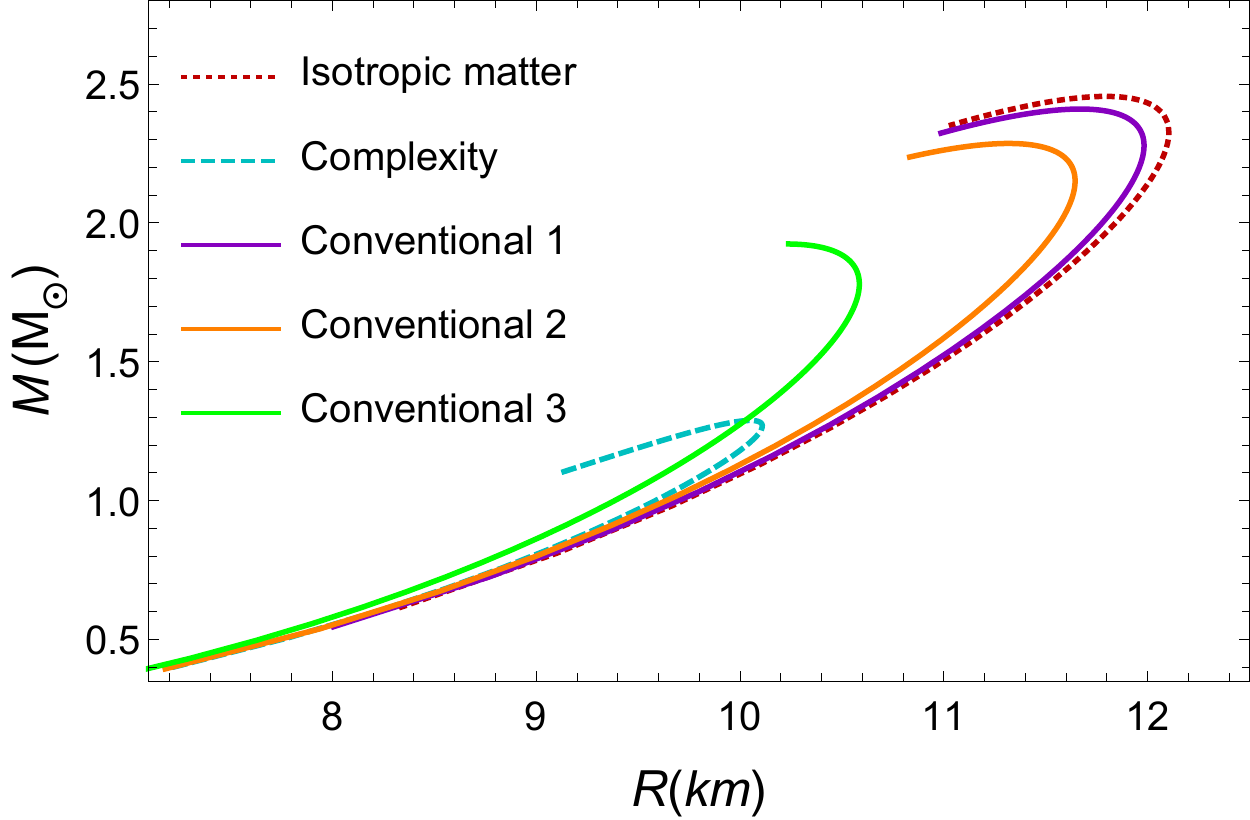} 
\caption{
The  mass-to-radius profiles for Model 1 in different cases: the dashed curve represents the profile when considering isotropic matter; the cyan curve corresponds to utilizing the vanishing complexity factor formalism; and the three curves in between depict the conventional method for values of $a$ equal to $100\,km$ (less anisotropic), $50\,km$ (moderate anisotropy), and $25\,km$ (more anisotropic).
}
\label{fig:7} 	
\end{figure*}


\section{Conclusions}

To summarize our work, we have obtained interior solutions of exotic stars made of dark energy, taking into account the presence of anisotropies and adopting the extended Chaplygin gas equation-of-state. The anisotropic factor is treated employing the formalism based on the complexity factor, and the structure equations have been integrated numerically. The solutions are shown to be well-behaved and realistic. Moreover, we have made a comparison with another more conventional approach, where the form of the anisotropic factor is introduced by hand.

\section{Acknowledgments}

A.~R. is funded by the Mar{\'i}a Zambrano contract ZAMBRANO 21-25 (Spain). I.~L. thanks the Funda\c c\~ao para a Ci\^encia e Tecnologia (FCT), Portugal, for the financial support to the Center for Astrophysics and Gravitation (CENTRA/IST/ULisboa)  through the Grant Project~No.\\~UIDB/00099/2020  and Grant No. PTDC/FIS-AST/28920\\/2017. 


\bibliographystyle{unsrt}         
\bibliography{biblio_1}

\end{document}